\documentclass[aps,prl,reprint,showpacs,amsmath,amssymb]{revtex4-1}
\usepackage{graphicx}
\usepackage{amsmath,amssymb,bm}
\usepackage{subfigure}

\begin{document}

 \title{Effect of electron-phonon coupling on transmission through Luttinger liquid  hybridized with resonant level}

\author{Alexey Galda, Igor V.\ Yurkevich and Igor V.\ Lerner}
\affiliation{School of Physics and Astronomy, University of
Birmingham, Birmingham B15 2TT, United Kingdom}

\date{\today}
\abstract{
We show that electron-phonon coupling strongly affects transport properties of the Luttinger liquid hybridized with a resonant level. Namely, this coupling significantly modifies the effective energy-dependent width of the resonant level in two different geometries, corresponding to the resonant or antiresonant transmission in the Fermi gas. This leads to a rich phase diagram for a  metal-insulator transition induced by the hybridization with the resonant level.
}
 }
 \bibliographystyle{eplbib}

\newcommand\ii{\mathrm{i}}
\newcommand\ee{\mathrm{e}}
\newcommand\vF{v_{_{\rm F}}}
\newcommand\eF{\varepsilon _{_{\rm F}}}
\newcommand\bF{\beta_{_{\rm F}}}
\newcommand\sgn{\operatorname{sgn}}

\maketitle

Low-temperature electronic properties of one-dimensional (1D) systems (like quantum wires or nanotubes)  are strongly affected by electron-electron interactions. Electrons in
such  systems form  a Luttinger Liquid (LL) \cite{Tom:50}\nocite{Lutt:63,HALDANE:81,vDSh:98}. An arbitrarily weak repulsion in a clean  LL leads to power-law decay of various correlation functions with  exponents depending on the interaction strength. Such a decay which is a characteristic property of the LL   has been been experimentally observed in carbon nanotubes \cite{Bockrath:99}\nocite{Yao:99,Ishii:03,Lee:04} and various quantum wires \cite{Auslaender:02}\nocite{Slot:04,Levy:06,Kim:06} (see Ref.~\cite{1Dreview:10} for a recent review).

Inserting a potential impurity or a weak link (e.g., a tunnel barrier) into the LL results in a power-law suppression of  a local density of states (LDoS) at the impurity site   and  thus a suppression of the conductance at low temperatures $T$ \cite{KaneFis:92a,MatYueGlaz:93},
changes in  characteristics of the Fermi-edge singularity \cite{furusaki:97}  and Friedel oscillations \cite{EggerGrabert:95}\nocite{Affleck:02}, etc. If the barrier interrupting the LL  carries a discrete localized state resonant with the electron Fermi energy, its hybridization  with the electronic states in the leads results in a sharp resonant transmission  \cite{KaneFis:92b}\nocite{PolGorn:03,FurMatv:02,NazGlaz:03}. Similar to the Fermi liquid, it is described by the Breit--Wigner formula but with the resonance width $\Gamma_0$ replaced by an energy-dependent effective  width $\Gamma({\varepsilon })$ vanishing at the Fermi level. Such a resonant transmission can be realized, e.g., by inserting into a 1D quantum wire a double barrier with a resonant level or a weakly coupled quantum dot (QD) with sufficiently large level spacing $\delta$ ($\delta\ll T,\Gamma_0$) and one level in resonance. We would refer to this geometry as resonant-barrier.

In a dual geometry, when a QD with such a resonant level is side-attached to the LL, transmission becomes antiresonant: it is reflectance rather than transmittance which is described by the   Breit-Wigner formula but with the width $\Gamma({\varepsilon })$  being power-law \textit{divergent} at the Fermi level \cite{LYY:08}\nocite{GB:10} (with the divergence cut by a temperature $T$). Both geometries are realistic: transmission and tunnelling measurements in the presence of controlled defects   have already been performed in both quantum wires \cite{Auslaender:02}  and  carbon nanotubes for various types of defects \cite{Bockrath:01}. Naturally in any realistic geometry electrons in the LL are inevitably interact not only with each other but also with phonons.

It is known that the electron-phonon (el-ph) coupling in combination with the electron-electron (el-el) repulsion  results in the formation of two polaron branches with different propagation velocities in the clean LL (see, for example,  \cite{Loss94,HoCaz}\nocite{F-BmixLutt}). The formation of polarons modifies the values of exponents in power laws characteristic for the LL. The exponents can change signs as functions of the relative strength of the el-el and el-ph coupling and of the ratio of the Fermi to sound velocities. The effect  of this is especially pronounced for the LL with an embedded potential scatterer. While a single scatterer embedded into the phononless LL makes it going from an ideal metal to an ideal insulator (at $T=0$) \cite{KaneFis:92a}, in the presence of the el-ph coupling such a transition can be reversed for a weak scatterer \cite{martin05} or, in general, can become dependent on the scatterer strength \cite{GYL:10}.

In this Letter we  show that the el-ph coupling results also in a drastic change of electronic transport trough the LL hybridized with a resonant level both in the resonant-barrier and in the side-attached geometry (see fig. 1).
\begin{figure}  \includegraphics[width=\columnwidth]{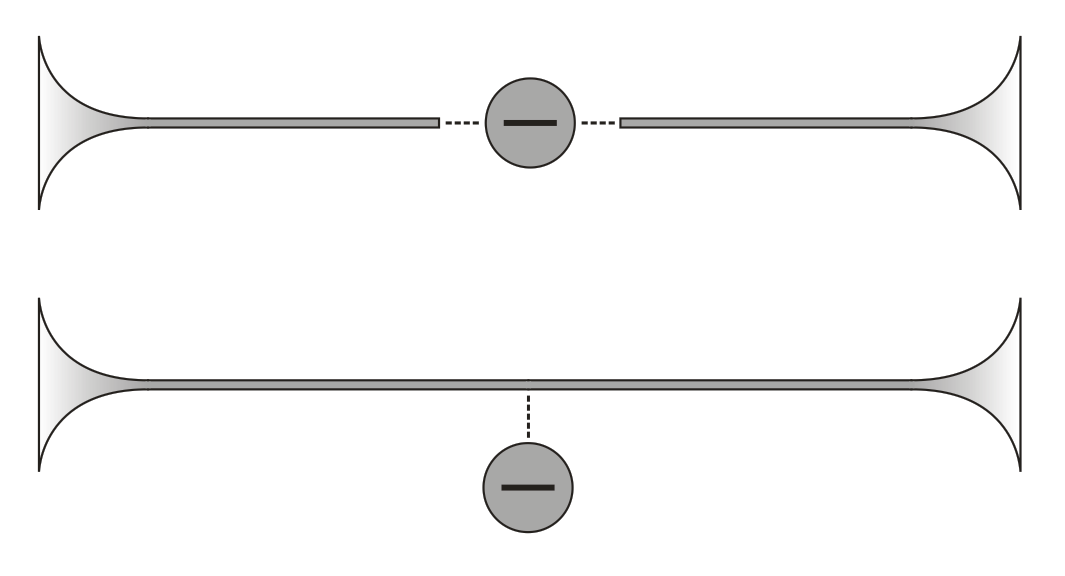}\\
  \caption{The geometries considered: resonant-barrier geometry (top) and side-attached geometry (bottom), also referred to in the text as RBG and SAG.}
\end{figure}

We will proceed as follows. First we introduce the action for the LL with the Coulomb repulsion,  el-ph coupling and hybridization with a resonant level. By integrating out  the phonon fields and the fields describing the resonant-level electron, we  obtain an effective action in terms of only the fields describing  the conduction electrons. Then we employ the functional bosonization  in form developed in \cite{GYL:04} to describe  the polaron formation in the presence of the resonant level in terms of the mixed fermion-bosonic action. Finally we  use the renormalization group (RG) analysis in form similar to that for the phononless LL \cite{KaneFis:92a,LYY:08} to calculate transmission through such a polaronic liquid as well as the effect of the el-ph coupling and  hybridization on the electronic LDoS in the vicinity of the QD.

\section{Effective action}
The action consists of three terms, $
S=S_\mathrm{ LL}+S_\mathrm{ el-ph}+S_\mathrm{ T}$.
We assume the usual LL decoupling of the (spinless) electron field   into the sum of right- and left-moving electrons,
\begin{align}\notag
    \psi({\xi })&=\psi_\mathrm{R}({\xi })\ee ^{\ii p_{_{\rm F}}x}+\psi_\mathrm{L} ({\xi })\ee ^{-\ii p_{_{\rm F}}x}\,,&\xi \equiv ({x,t}),
\end{align}
with   $\psi_\mathrm{R,L}$ being labelled by $\eta=\pm1$ below, and unit with ${\hbar}=1$ are used.   Then the LL part of the action, which includes the electron  kinetic energy and the density-density interaction, has the following form:
\begin{align}\label{LL}
{S}_{\rm LL}&= \!\! \!\sum_{\eta=\pm1} \int\!\!\ii  \xi\,{\bar\psi}_{\eta}({\xi })\, \ii \partial_{\eta} \psi_{\eta}({\xi })-\frac{1}{2}  \int\!\!\ii  \xi \,V_0 \,n^2 ({\xi })\,.
\end{align}
Here  $n\equiv ({{\bar\psi}_\mathrm{R}\psi_\mathrm{R}+ {\bar\psi}_\mathrm{L}\psi_\mathrm{L}})$ is the electron density, $V_0$ is the screened Coulomb interaction, $\ii \xi $ stands for $\ii  x\,\ii  t$ and $\partial_{\eta} \equiv \partial_{t}+\eta \vF \partial_{x}$.  The free conduction electron Green function, $g_\eta({\xi-\xi' })$, is defined by $\partial_{\eta} g_\eta({\xi-\xi ' })=\delta({\xi -\xi ' })$ so that its Fourier transform is given by
\begin{align}\label{FGF}
g_\eta (\varepsilon,\,q)&=\left[ \varepsilon +\ii\delta\sgn {\varepsilon} -\eta \vF  q\right]^{-1}\,.
\end{align}
We use here the zero-temperature formalism: the only role played by temperature is providing an alternative low-energy cutoff in the RG equations below.

It is not particularly important for what follows how we model the phonon action. We choose a model of 1D acoustic phonons linearly coupled to the electron density, assuming the phonon spectrum to be linear with a cutoff at the Debye frequency, $\omega_{_{\rm D}}=cq_{_{\rm D}}$. Then the phonon action (neglecting possible electron backscattering   which is justified for  $T\ll\omega_D$) has the following form:
\begin{equation}\label{el-ph}
S_\mathrm{el-ph}=\int\!\!\ii  \xi\Bigl[-\frac{1}{2}\phi({\xi })\, \mathcal{D}_{0}^{-1}\,\phi({\xi })+g\,\phi({\xi })\,n({\xi })\Bigr]\,.
\end{equation}
Here $\phi ({\xi })$ is the phonon field,  $g$ the el-ph coupling constant and  $\mathcal{D}_0({\xi })$ the free phonon propagator with the Fourier transform given by
\begin{align}\notag
\mathcal{D}_0({\omega,\,q})& =\frac{   \omega_q^2} { \omega  ^2-\omega_q^2+\ii\delta}\,,&\omega_q&=cq\,.
\end{align}

Finally, the tunnelling action for both  geometries (assuming that the impurity  or QD  carrying the resonant level is inserted in the LL or side-attached to it at $x=0$) has the  form:
\begin{align}\label{ST}
S_\mathrm{T}=\!\int\!\ii  t\Big\{\ii{\bar d}(t)\left(\partial_{t} +\varepsilon _0 \right)d(t)+\sum_{\mu} \big[ t_0{\bar d}({t})\psi_{\mu}({t})+ \mathrm{h.c.}\big]\Big\}\,,
\end{align}
where $\psi_{\mu}({t})\equiv\psi_{\mu}(x=0,{t})$,  $d({t})$ is the field corresponding to the electron localized at the resonant level with the energy $\varepsilon _0$  counted from the Fermi level. In the side-attached geometry (SAG) the index $\mu$ simply labels right- and left-movers while in the resonant-barriers geometry (RBG) it refers to the left and right electron subsystems separated by the barrier. In this case the electron can leave the left subsystem to the QD and enter it from the QD only as a right- and left-mover, respectively -- and conversely for the right subsystem so that the labels $\mu=\pm1$ mean
 \begin{align}\label{psi}
 \bar{\psi}_-=\bar{\psi}_{\ell,L}\,,\; \psi_-=\psi_{\ell ,R} \,,\;\bar{\psi}_+=\bar{\psi}_{r,R}\,,\; \psi_+=\psi_{r,L}  \,
\end{align}
where $L,\,R$ refer to the left- and right movers, as before, and $\ell ,\, r$ to the left and right subsystems. We have assumed the tunnelling amplitude $t_0$ to be the same in all channels. This is always the case for the SAG, but not necessarily for the RBG if it is implemented as an asymmetric double-barrier. In the latter case the resonant properties could be different for $t_1\ne t_2$ but we leave it aside here.

The action defined by Eqs.~(\ref{LL}), (\ref{el-ph}) and (\ref{ST}) is quadratic in fields $d$ and $\phi $ which can thus be integrated out. The integration over the phonon fields results only in substituting $V_0$ in the action (\ref{LL}) by the dynamical coupling,
\begin{align}\label{V}
    V({\xi })=V_0+g^2\mathcal{D}_0 (\xi )\,.
\end{align}
Performing the integration over the field $ d(t)$ results in transforming the tunnelling term (\ref{ST}) in the action to the following one:
\begin{align}\notag
\tilde{S}_\textrm{T}&=-\sum_{\mu,\nu}\! \int\!\!\ii  t  \,\ii  t'  {\, \bar\psi}_{\mu}({t  })\,\Sigma_{\mu,\nu} ({t -t' })\,\psi_{\nu}({t'  })\,,
\\[-8pt]\label{sigma0}\\[-8pt]
\notag\hat\Sigma({\varepsilon }) &=\frac{ \vF \hat\Gamma_0}{ {{\varepsilon -\varepsilon _0 +\ii\delta\sgn {\varepsilon} }}  }\,,\qquad
  \hat\Gamma_0 \equiv \Gamma_0({\hat{1}+\hat{\sigma_x}})\,,
\end{align}
where  $\Sigma({t-t'})$ is the Fourier transform of $\Sigma({\varepsilon })$. $\hat\Gamma_0$ is the matrix with all elements equal to the tunnelling rate $\Gamma_0\equiv\pi\nu_0|t_0|^2$, where $\nu_0=({\pi \vF})^{-1}$ is the one-particle DoS of the  conduction electrons in the absence of interactions.

The action given by Eqs.~(\ref{LL}) (with $V_0=0$) and (\ref{sigma0}) describes the resonant transmission through the Fermi gas hybridized with the resonant level. The hybridization makes the electron Green function to acquire an off-diagonal part, $g_\mu(\xi -\xi ')\to G_{\mu\nu}(\xi ,\xi ')$, describing  the resonance-induced backscattering in the SAG or left-to-right connection in the RBG. In the mixed position-energy representation it has the following matrix form
\begin{align}\label{HGF}
   \hat{G} (x,x';\varepsilon)&=\hat{g}
(x-x';\varepsilon) +  \ii\vF\hat{g}
(x;\varepsilon) \hat{\mathsf{T}}({\varepsilon })\,\hat{g}(-x';\varepsilon) \, .
\end{align}
Here  $\hat{g}$ is the  matrix with  diagonal  elements given by eq.~(\ref{FGF}), and the  $\hat{\mathsf{T}}$-matrix  has the form
\begin{align}\label{TN}
  \hat {\mathsf{T}}({\varepsilon })&=
  \frac{-\ii\hat\Gamma_0} { {\varepsilon -\varepsilon _0+\frac\ii2 \hat\Gamma_0\sgn\varepsilon } }\,.
\end{align}
Although the Green function (\ref{HGF}) is formally the same for both geometries considered, the  transmission for the RBG is proportional to $G_{12}G_{21}$ while for the SAG it is proportional to $G_{11}G_{22}$. This gives the well-known Fermi-gas result with the resonant transmission for the RBG and the resonant reflection for the SAG (with $\mathcal{R}_0=1-\mathcal{T}_0$):
\begin{align}\label{TP}
    \mathcal{T}_0(\varepsilon)  =\left\{\begin{array}{cl}
\dfrac{\Gamma_0 ^2} {(\varepsilon -
\varepsilon_0)^2 + \Gamma^2_0} \,, & \text{RBG} \\[9pt]
\dfrac{(\varepsilon - \varepsilon_0)^2} {(\varepsilon -
\varepsilon_0)^2 + \Gamma^2_0} \,, & \text{SAG}\;.
                                        \end{array}
    \right.
\end{align}

We will use the effective action   represented by the sum of the  terms given by eq.~(\ref{LL}) with the substitution (\ref{V}) and eq.~(\ref{sigma0}) to show that  the el-el and el-ph interactions results in $\Gamma_0\to\Gamma({\varepsilon })$ in the transmission probability (\ref{TP}), with the energy dependence of $\Gamma({\varepsilon })$ being qualitatively different from that found in the phononless case \cite{KaneFis:92b,LYY:08}.

\section{Functional bosonization}
The first step is the Hubbard--Stratonovich transformation which decouples the $n^2$ term in the action (\ref{LL}) with the substitution (\ref{V})  and results in the mixed fermionic-bosonic action in terms of the auxiliary bosonic field $\varphi$ minimally coupled to $\psi$:
\begin{equation}\label{S}
S_{\rm eff}=-\frac{1}{2}\int\!\!\ii  \xi\,\varphi\,V^{-1}\,\varphi +\ii \!\! \sum_{\eta=\pm1} \int\!\!\ii  \xi\, {\bar\psi}_{\eta}\left(\partial_{\eta}-\varphi\right)\psi_{\eta}\,.
\end{equation}
Here $\eta$ labels right- and left-moving electrons for both geometries under consideration. We stress that for the RBG  there is no interaction between electrons in different halves of the system and in this case the action (\ref{S}) describes electrons moving in one of the subsystems which are connected only via the resonant tunnelling. We will refer to  electrons in different ($\ell $ and $r$) subsystems using indices $\mu,\nu=\pm1$ which simultaneously label right- and left-movers as in eq.~(\ref{psi}), keeping $\eta,\eta'$ for referring only to the right- and left-movers in both geometries.

The introduction of the field $\varphi $ in eq.~(\ref{S}) does not constitute the functional bosonization: the latter is in getting rid of the coupling term by the following gauge transformation which introduces the new bosonic field $\theta({\xi })\equiv\theta({x,t})$:
 \begin{align}\label{gauge}
\psi_{\eta}({\xi }) &\to \ee^{\ii\theta_{\eta}({\xi }) }\psi_{\eta}({\xi }) \,,& \ii\partial_{\eta} \theta_\eta({\xi }) &=\varphi ({\xi })\,.
\end{align}
The Jacobian of this transformation results \cite{GYL:04} in substituting $  V^{-1}+\Pi$ for $V^{-1}$ in  eq.~(\ref{S}), where $\Pi=\Pi_\mathrm{R} +\Pi_ \mathrm{L}$ is the  one-loop electronic polarization operator (exact for the LL), with $\Pi_\eta(\xi )=\ii g_\eta({\xi })g_\eta({-\xi }) $ and $g_\eta({\xi })$ given by eq.~(\ref{FGF}). Note that one could have started with the unscreened Coulomb interaction, as this transformation provides the screening which is, naturally, identical to one originally calculated diagrammatically  \cite{DzyalLar:73}.

The conduction electron Green function is not invariant under the gauge transform (\ref{gauge}) and so it is dressed by $\theta({\xi })$. In the absence of the resonance part of the action,  eq.~(\ref{sigma0}), the dressed Green function for right- or left-moving electrons is given by
\begin{equation*}
G_{\eta}(\xi - \xi ')=g_{\eta}(\xi , \xi ')\,\ee^{\ii U_{\eta\eta}(\xi ,\xi ')}\,,
\end{equation*}
where $\ii U_{\eta\eta'}\equiv\langle\theta_{\eta}\theta_{\eta'}\rangle$ is the correlation function of the fields $\theta$. To calculate it we notice that eq.~(\ref{gauge}) for the gauge field  $\theta({\xi })$  is resolved with the help of the bosonic Green function $g_{\eta}$  which coincides with the free electron Green function (\ref{FGF}):
\begin{equation*}
\theta_{\eta}(\xi )=\int \!\!\ii \xi '\,g _{\eta}(\xi -\xi ')\,\varphi(\xi ')\,,
\end{equation*}
Thus we obtain the Fourier transform of $U_{\eta\eta'}$ as follows:
\begin{equation}\label{U}
U_{\eta\eta'} (\omega, q)= \frac{\omega_+\!+\!\eta'\vF q}{\omega_+\!-\!\eta\phantom' \vF q}\,
\frac{V_0 (\omega_+^2-\omega^2_q) +g^2\omega^2_q} {\left(\omega_+^2-v^2_+q^2\right)\left(\omega_+^2-v^2_-q^2\right)},
\end{equation}
where $\omega_+\equiv\omega+\ii\delta\sgn\omega$ and $v_{\pm}$ are  velocities of the composite bosonic modes (polarons) given by
\begin{equation}\label{pm}
v^2_{\pm}=\frac{1}{2}\left[v^2+c^2\pm\sqrt{\left(v^2-c^2\right)^2 + 4\alpha_\mathrm{ph}  \vF^2c^2}\,\right]\,.
\end{equation}
Here we introduced the dimensional el-ph coupling constant $\alpha_\mathrm{ph} \equiv \nu_0 g^2$ while $v$ is the speed of plasmonic excitations in the phononless  LL,
$$   v =\vF ({1+\nu_0V_0})^{1/2}\equiv \vF K^{-1},
$$
where $K $ is the standard Luttinger parameter. We assume that in the absence of phonons the el-el interaction is always repulsive ($V_0>0$) so that $K<1$. Note that in the limit corresponding to the absence of phonons  ($c=0$), one has $v_-=0$ and $v_+ = v$, so that in this case $U_{\mu\nu}$ reduces to the usual LL plasmonic propagator. This also happens in the absence of the el-ph coupling ($\alpha_\mathrm{ph} =0$) when $v_-=\min(c,v) $ and $v_+=\max(c,v)$ and these two branches are totally decoupled. We take the same limit for $U$ when  $\omega>\omega_{_{\rm D}}$ so that $c$ should be put to $0$.

Equations (\ref{U}) and (\ref{pm}) describe well known two-branch polaronic excitations \cite{Loss94} in the LL in the presence of the  el-ph coupling. The slow and fast branches have the velocity $v_\mp$ obeying the inequalities $v_- < c,\,v < v_+$. Here we have restricted our considerations to the stability region defined by
\begin{align}\label{alpha}
    \alpha\equiv\alpha_\mathrm{ph}K^2<1\,,
\end{align}
where $v_-^2>0$, i.e.\ we leave out of considerations the Wentzel--Bardeen instability \cite{Wentzel}\nocite{Bardeen:51}.

The existence of the two polaron branches may be interpreted as splitting the LL in the presence of the el-ph coupling into the two-component liquid, with the effective Luttinger parameters
$
K_{\rm fast}={\vF }/{v_+}<K<1\,,\quad K_{\rm slow}={\vF }/{v_-}>1,
$
corresponding to el-el repulsion (which becomes stronger with the el-ph coupling) and the phonon-mediated attraction. The two-mode nature of the el-ph LL drastically changes a character of the resonant (or antiresonant) transmission.

The self-energy part  in resonant-level term (\ref{sigma0}) is dressed as a result of the gauge transform (\ref{gauge}) by the local field $\theta({\varepsilon })\equiv \theta({x=0,t})$:
\begin{equation}\label{sigma}
\Sigma ({t-t'})\to \Sigma_{\mu\nu}({t-t'})= \ee^{-\ii\theta_{\mu}({t})}\,\Sigma ({t-t'})\,\ee^{\ii\theta_{\nu}({t'})}.
\end{equation}
It is this dressing which fully governs the resonance-width renormalization, $\Gamma_0\to\Gamma({\varepsilon })$, in the presence of the el-el and el-ph interactions.

\section{The self-energy renormalization}
The polaron fields $\theta_{\mu}({t})$ entering the self-energy (\ref{sigma}) are defined at the origin where the QD (or barriers) carrying the resonant level is placed. Integrating out all the fields at $x\ne 0$ results in the zero-dimensional action which governs the renormalization of the self-energy in Eqs.~(\ref{sigma0}) and (\ref{sigma}), and thus the renormalization of the resonance width in Eqs.~(\ref{TP}). In the phononless case this action is fully equivalent to that used for describing the resonant transmission (reflection) through the LL \cite{KaneFis:92b,LYY:08}. The only difference due to the el-ph coupling is that the  correlation function of the local fields $\theta({t})$ is governed by the two-branch polaron modes, eq.~(\ref{U}). It  follows from this equation that
\begin{align}\label{u}
\langle\theta_{\mu}(-\omega)\,\theta_{\nu}(\omega)\rangle =\ii\int \!\frac{\ii  q}{2\pi}  \,U_{\mu\nu}({\omega; q})\equiv \frac{\pi\gamma_{\mu\nu}}{|\omega|}\,,
\end{align}
where the dimensionless correlation matrix $\gamma_{\mu\nu}$  found from  the  above integration can be parameterized as
\begin{equation}\label{gamma}
\gamma_{\mu\nu}=\left\{\begin{array}{ll}
                         \delta_{\mu\nu}\,\gamma_+\,, & \text{ RBG }  \\[8pt]
                         \frac{1}{2}\left(\gamma_+-\gamma_-\right)+\gamma_-\, \delta_{\mu\nu}\,, & \text{ SAG\,. }
                       \end{array}
 \right.
\end{equation}
Here $\gamma_\pm$ can be represented as
\begin{align}\label{gpm}
\gamma_{+}&=\frac{\kappa_+}{K}-1\,,& \gamma_{-}&=\kappa_{-}K-1\,,
\end{align}
where one would have $\kappa_{\pm}=1$ without coupling to phonons in which case $\gamma_+\to\tilde{\gamma}_+\equiv1/K-1$, the exponent describing renormalization of the conductance by a weak scatterer, and $\gamma_-\to\tilde{\gamma}_-\equiv K-1$, the exponent describing renormalization  by a weak link. The el-ph coupling modifies these exponents by the factors $\kappa_{\pm}(K)$ given by
\begin{subequations}\label{kappa} \begin{align}
\kappa_{+}&= \left[1+\frac{\alpha}{\left({\beta+\sqrt{1-\alpha} }\right)^2} \right]^{-1/2}\!, \qquad \beta\equiv\frac{v}{c}\,,
 \\ \kappa_{-}&= \left\{\left(1-\alpha\right)\left[1+\frac{\alpha}{\left({\beta^{-1}+\sqrt{1-\alpha} }\right)^2} \right]\right\}^{-1/2}\!.
\end{align} \end{subequations}

It is easy to verify that $\kappa_+(K)\leq 1$ while $\kappa_-({K})\geq1$, so that $\gamma_+<\tilde{\gamma}_+$ and $\gamma_->\tilde{\gamma}_-$. Therefore, whereas the phononless exponents are sign-definite, $\tilde{\gamma}_+>0$ and $\tilde{\gamma}_-<0$, this is not necessarily true for the exponents $\gamma_+$ and $\gamma_-$. We have previously shown \cite{GYL:10} that electronic transport through the LL with a weak scatter or with a weak link, described by the correlation functions with the exponents $\gamma_\pm$ respectively, is strongly influenced by the el-ph coupling due to these exponents changing sign at different values of parameters $\alpha$ and $\beta$. Here we will show that this also strongly affects the resonant transmission (reflection) in both geometries  where the exponents $\gamma_\pm$ are changed by the el-ph interaction enter via eq.~(\ref{gamma}).

As in the case of the resonance transport through the phononless LL  \cite{KaneFis:92b,LYY:08}, we shall write a renormalization group (RG) equation for the tunnelling amplitude $t_0$. In our formalism we should start with renormalizing the self-energy part $\Sigma$ in the tunnelling action (\ref{sigma0}) with the gauge substitution (\ref{sigma}).

Having integrated out  the fields with $x\ne0$, all the interaction effects enter via the correlation functions of the bosonic field $\theta$, defined by Eqs.~(\ref{u}) -- (\ref{kappa}). Therefore, the RG equation for $\Sigma$ is obtained by  a usual integration  over fast components of this field. We do not integrate over  fast components of the fermionic field $\psi({t})$ and do not rescale the time variable since these two procedures exactly cancel each other which follows from the absence of the renormalization of the self-energy in the noninteracting case.  We assume that the fast Fourier components of the  field $\theta_{\mu}({t})$ have frequencies $ E\leq|\omega|\leq E'$, with $E$ being the running cutoff and $E'/E-1\ll1$.  Integrating them out leads to the following increment for the self-energy:
\begin{align}\notag
\delta\Sigma_{\mu\nu}(\varepsilon)=-\!\!\!\!\!\int\limits_{ E\leq|\omega| \leq E' } \frac{\mathrm{ d}\omega}{2|\omega|} \biggl[&\frac{\gamma_{\mu\mu}+\gamma_{\nu\nu}}{2} \Sigma_{\mu\nu} (\varepsilon)\\&\label{IRG}-\gamma_{\mu\nu} \Sigma_{\mu\nu}(\varepsilon+\omega)\biggr]\,.
\end{align}
In such an RG scheme \cite{KaneFis:92b} the self-energy $\Sigma_{\mu\nu}$ acquires a dependence on the running cutoff $E$ on top of the dependence on $\varepsilon $. The initial condition for the RG equations is  that at the ultraviolet cutoff, $E=E_0$, ($E_0\sim \eF$ is the bandwidth), $\Sigma_{\mu\nu}$ is independent of $E$ and has all matrix   elements equal to $\Sigma({\varepsilon })  $  given by eq.~(\ref{sigma0}). Then as long as $E\gg|\varepsilon -\varepsilon _0|$ one may discard the second term in eq.~(\ref{IRG}), thus arriving at the following RG equation:
\begin{align}
 \label{RG1}
\frac{\mathrm{ d}\Sigma_{\mu\nu}(\varepsilon;E)}{\mathrm{ d}l}&=-\gamma_{\mathrm{d}}\Sigma_{\mu\nu}(\varepsilon;E)\,,&
\end{align}
where $l\equiv\ln E_0/E$ and $\gamma_{\mathrm{d}}$ are  equal diagonal components  of the matrix $\gamma_{\mu\nu}$, eq.~(\ref{gamma}), given by
\begin{align}\label{gd}
    \gamma_{\mathrm{d}}=
\left\{\begin{array}{cl}
         \gamma_+\, ,& {\text{RBG}}  \,,\\
         \gamma_0\equiv \frac1{2}({\gamma_++\gamma_-})\, ,&  {\text{SAG}}  \,.
       \end{array}
\right.
\end{align}
Note that $\gamma_+$ and $\gamma_0$
happen to be the edge and bulk DoS exponents respectively, equal to $1/K-1$ and $(1-K)^2/2K$ in the phononless case \cite{KaneFis:92a} and given by eq.~(\ref{gamma}) and (\ref{kappa}) in the presence of the el-ph interaction.

\begin{figure}  \subfigure[Resonant-barrier geometry: $ \Gamma_{\mathrm{off}}\propto |\varepsilon /E_0|^{\gamma_{+} }$ with the sign of $\gamma_+$  depending on the  el-ph coupling strength. $\Gamma_{\mathrm{diag}}=\Gamma_{\mathrm{off}}$ at $\varepsilon \gtrsim \varepsilon _0$ and saturates at $\varepsilon_0$ ({dashed line}) for  $\varepsilon \lesssim\varepsilon_0$. ] {\includegraphics[width=.9\columnwidth
]{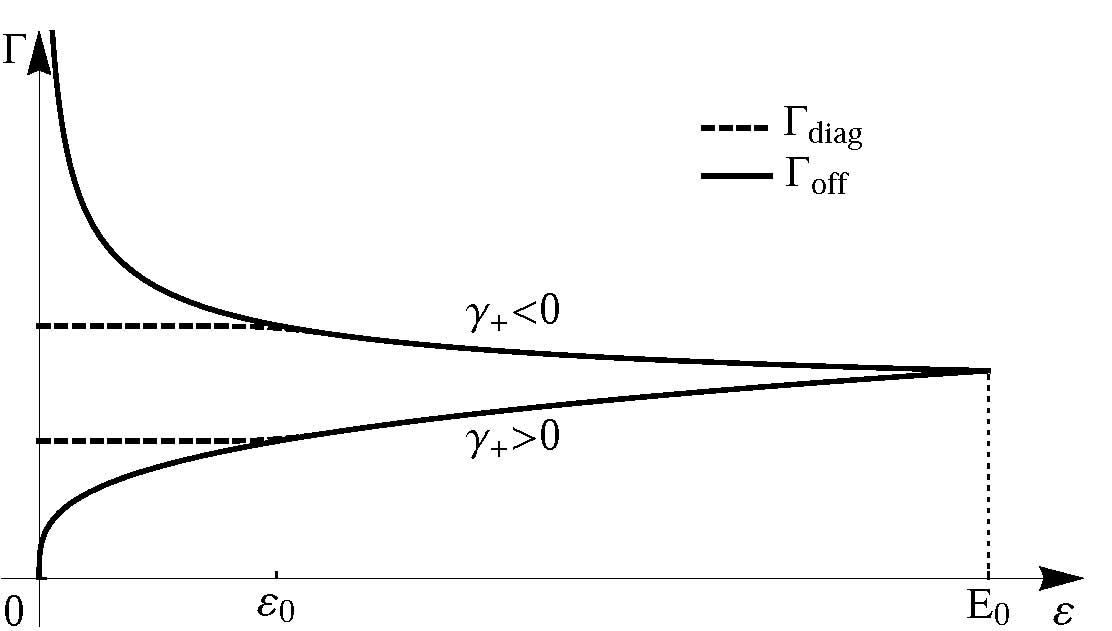}}
\subfigure[Side-attached geometry: $\Gamma_{\mathrm{diag}}=\Gamma_{\mathrm{off}}\propto |\varepsilon /E_0|^{\gamma_{0} }$ for \mbox{$\varepsilon \gtrsim \varepsilon _0$};  $\Gamma_{\mathrm{diag}}$ saturates  and $\Gamma_{\mathrm{off}}\propto |\varepsilon /\varepsilon _0|^{\gamma_-}$  for $\varepsilon \lesssim\varepsilon_0$. The sign of $\gamma_-$  depends on the  el-ph coupling strength, while $\gamma_0>0.$
 ] {\includegraphics[width=.9\columnwidth
]{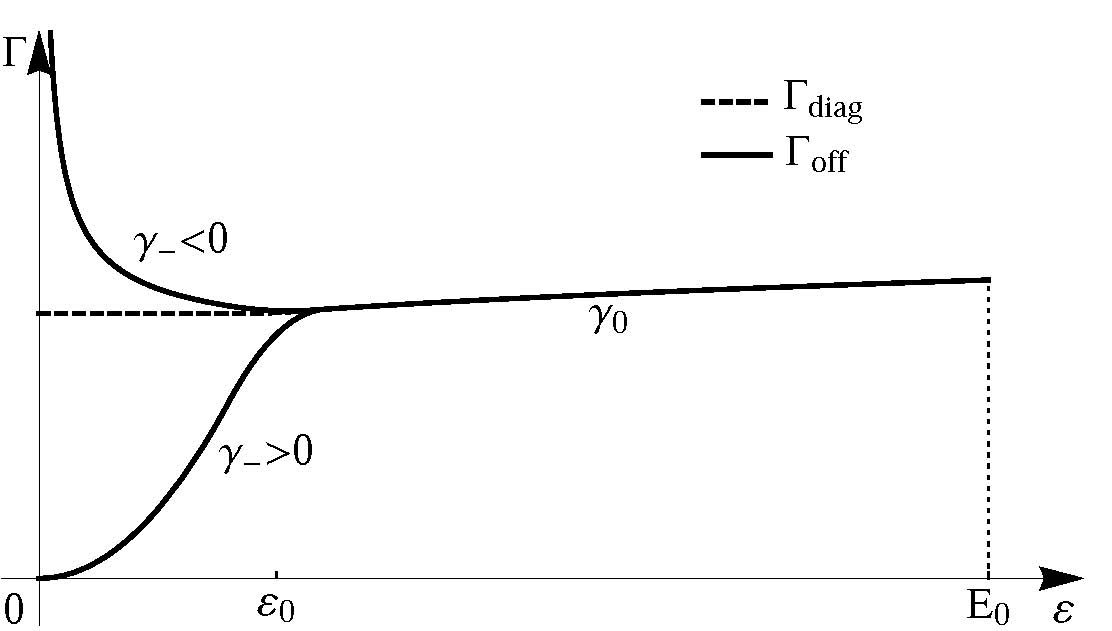}}
\caption{\label{GG} $\Gamma_{\mathrm{diag}}({\varepsilon }) $ and $\Gamma_{\mathrm{off}}({\varepsilon }) $ for  RBG   and SAG (not to scale).   }\label{Gamma}\end{figure}

Equation (\ref{RG1}) is solved by  substituting
$
\Sigma_{\mu\nu}(\varepsilon;E)$ in form (\ref{sigma0}), i.e.\ with all matrix elements equal, but with $\Gamma_0$ replaced by $\Gamma({E})$. This leads to   the  RG equation for $\Gamma$:
\begin{align}\label{t}
\frac{\mathrm{ d}\Gamma (E)}{\mathrm{ d}l}&=-
  \gamma_{{\mathrm{d}} }\,\Gamma (E)\,,  & E&\gg|\varepsilon -\varepsilon _0|\,.
\end{align}
With lowering the running cutoff one eventually reaches the region $E\ll|\varepsilon-\varepsilon_0|$  where the second term must be taken into account. The self-energy still has the form of eq.~(\ref{sigma0}) but with the substitution
\begin{align}\label{subs}
\hat{\Gamma}_0\to\hat\Gamma({E})=\Gamma_{\mathrm{diag}} ({E})\hat{1}+
 \Gamma_{\mathrm{off}}({E})\hat\sigma_x\,.
\end{align}
For $E\ll|\varepsilon-\varepsilon_0|$   the second term in eq.~(\ref{IRG})  cancels the first one for $\mu=\nu$  so that  $\Gamma_{\mathrm{diag}} $ saturates at $\Gamma(|\varepsilon -\varepsilon _0|)$ obtained by solving eq.~(\ref{t}), while $\Gamma_{\mathrm{off}} $ continues to be renormalized according to the following RG equation:
\begin{align}\label{t2}
    \frac{{\mathrm{d} \Gamma_{\mathrm{off}}({E}) } }{{\mathrm{d}}l }&=-\gamma_\pm\Gamma_{\mathrm{off}}({E})\,, & E&\ll|\varepsilon -\min\{{\Gamma_0,\,\varepsilon _0}\}|\,,
\end{align}
with $\gamma_+$ for the RBG, as in eq.~(\ref{t}), and $\gamma_-$ for the SAG. Behavior of renormalized $\Gamma({\varepsilon })$ is shown in fig.~\ref{GG}. Let us stress that the condition of  applicability written above does not follow directly from  eq.~(\ref{IRG}) which is perturbative in $\Gamma_0$. However, it follows from considerations non-perturbative in tunnelling (but perturbative in the interaction strength) \cite{MatYueGlaz:93,LYY:08} that the inequality (\ref{t2}) actually means the off-resonance condition. The impurity remains off-resonant if the level width renormalized according to eq.~(\ref{t}) remains narrow, i.e.\ $\Gamma({\varepsilon _0})\ll \varepsilon _0$.  Only in this case $\Gamma_{{\mathrm{off}} } $ renormalizes as  in eq.~(\ref{t2}). Otherwise, we should put $\varepsilon _0=0$ and describe the resonant situation entirely in the frame of eq.~(\ref{t}).

\section{Transmission coefficient}
The  transmission coefficient, ${\cal T}(\varepsilon)$, is obtained    by replacing in the $\mathsf{\hat T}$-matrix   (\ref{TN}) the bare  tunneling rates, $\hat{\Gamma}_0$,  by  the renormalized ones, $\hat{\Gamma}(E\!=\!\varepsilon )$,   found from  Eqs.~(\ref{t}) and (\ref{t2}).
The off-diagonal element of the transmission matrix, $\mathsf{T}_{\mathrm{off}}$, is equal to the transmission or reflection amplitude for, respectively, the RBG or SAG  so that
\begin{align}
\label{T}
{\cal T}({\varepsilon })&=\left\{
\begin{array}{cl}
  |\mathsf{T}_{\mathrm{off}}({\varepsilon })|^2\,, & \mathrm{RBG} \\
  1-|\mathsf{T}_{\mathrm{off}}({\varepsilon })|^2\,, & \mathrm{SAG}
\end{array}\right.\,.
\end{align}
Here $|\mathsf{T}_{\mathrm{off}}({\varepsilon })|^2$ has the following form found  from eq.~(\ref{TN}) with the substitution (\ref{subs}) at $E={\varepsilon }$:
\begin{align}\label{Toff}
|\mathsf{T}_{\mathrm{off}}|^2&=\frac{\Gamma_{\mathrm{off}}^2} {\left(\varepsilon_0-\varepsilon+\Lambda\right)^2+\Gamma^2_{\mathrm{diag}}}\,,&
    \Lambda&\equiv \frac{\Gamma_{\mathrm{off}}^2-\Gamma^2_{\mathrm{diag}}}{4\varepsilon_0}\,.
\end{align}
For  $\varepsilon \gtrsim|\varepsilon_0|$, the diagonal and off-diagonal elements of ${\hat\Gamma}$ are equal and renormalize in the same way, eq.~(\ref{t}). This leads to the transmission coefficients of the Fermi-gas form (\ref{TP}) but with the renormalized tunneling rate,
\begin{align}\label{Thigh}
 \Gamma(\varepsilon)=\Gamma_0\left(\frac{\varepsilon}{E_0}\right)^{\!\gamma_{\mathrm{d}} }\,,
\end{align}
which fully describes the  case of resonance.
When the impurity level is off-resonance, we  substitute $\varepsilon _0$  for $\varepsilon_0-\varepsilon $ into eq.~(\ref{Toff})  and take into account that $\Gamma_{\mathrm{off}} $ continues to renormalize,
eq.~(\ref{t2}), while
$\Gamma_{\mathrm{diag}} $ saturates:
\begin{align}
    \Gamma_{\mathrm{off}}(\varepsilon)&=\Gamma(\varepsilon_0)
  \left(\frac\varepsilon{\varepsilon_0}\right)^{\gamma_\pm} \,,
  &
  {\Gamma}_{\mathrm{diag}} &=\Gamma (\varepsilon_0)\,.
\end{align}

\section{Resonant conductance} At nonzero but low temperatures $T$, the two-terminal conductance $g({T})$ is proportional to ${\mathcal{T}}({\varepsilon})$ with the low-energy cutoff at $\varepsilon \sim T$ (the Fermi energy corresponds to $\varepsilon =0$).  In the off-resonance situation, $|\mathsf{T}_{\mathrm{off}}({\varepsilon })|$ in eq.~(\ref{Toff})  vanishes with $\varepsilon \to0$: the resonant level remains decoupled from conduction electrons even when $\Gamma_{\mathrm{off}}({\varepsilon }) $ diverges when $\varepsilon \to0$.

\begin{figure}
\includegraphics[width=\columnwidth]{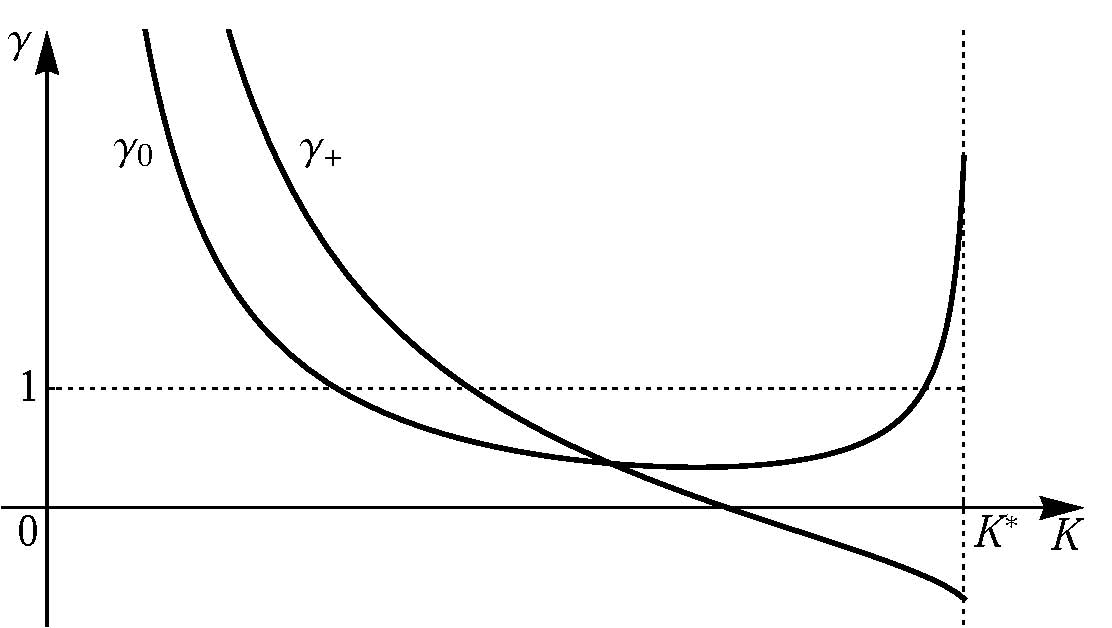}\\
  \caption{\label{Gdp} The RG exponents of the effective resonance width for the RBG, $\gamma_+$, and the SAG, $\gamma_0$. Here $K^*\equiv \min\{{1,\alpha_{\mathrm{ph}} }^{-\frac{1}{2}}\}$ is the boundary of the applicability region: for $\alpha_{\mathrm{ph}}>1 $ we stay away from the Wentzel--Bardeen instability \cite{Wentzel}. }
\end{figure}

The  resonance, $\varepsilon_0 \lesssim \Gamma({\varepsilon _0})$, is described by eq.~(\ref{Toff}) with $\Lambda=0$. This corresponds to the Fermi-gas expression, eq.~(\ref{TP}), with $\varepsilon _0=0$ and $\Gamma_0$ substituted by its renormalized value, $\Gamma({\varepsilon })$,  eq.~(\ref{Thigh}). The critical exponent $\gamma_{\mathrm{d}} $, eq.~(\ref{gd}), is strongly affected by the el-ph coupling, as illustrated in fig.~\ref{Gdp}: without phonons $\gamma_{\mathrm{d}}$ (i.e.\ $\gamma_+$ or $\gamma_0$) are monotonically  decreasing functions reaching $0$ at $K=1$.   Crucially, conductance $g$ is either ideal or vanishing at $T=0$, depending on   whether $\gamma_{\mathrm{d}} >1$ or $\gamma_{\mathrm{d}} <1$.

For $\gamma_{\mathrm{d}} >1$, $|\mathsf{T}_{\mathrm{off}}({\varepsilon })|$ in eq.~(\ref{Toff})  vanishes with $\varepsilon \to0$, as in the case of a strong el-el coupling  in the phononless LL ($K<1/2$ without pair tunnelling \cite{KaneFis:92a}).  This happens because $\Gamma({\varepsilon } ) \to0$ faster than $\varepsilon \to0$, i.e.\ the resonant level remains effectively decoupled from conduction electrons.

On the contrary, for $\gamma_{\mathrm{d}} <1$ we have $|\mathsf{T}_{\mathrm{off}}({\varepsilon })|\to1$  for $\varepsilon \to0$ which leads to an ideal resonance for the RBG and antiresonance for the SAG, eq.~(\ref{T}).

Thus the effective decoupling of the resonant electron level from conduction electrons at $\gamma_{\mathrm{d}} =1$ leads to a metal-insulator transition. Figure~\ref{Gdp} shows that for the RBG, where $\gamma_{\mathrm{d}} \equiv \gamma_+$, the el-ph coupling shifts the transition towards stronger el-el coupling. Note also that when $\gamma_+<0$, the effective resonance width diverges with $\varepsilon \to0$ rather than vanishes as in the phononless LL. For the SAG, where $\gamma_{\mathrm{d}} \equiv \gamma_0$, there could be two phase transitions: a sufficiently strong el-ph interaction can decouple the resonant level from conduction electrons, leading to the metallic phase, also for a weak el-el coupling when $K$ is close to $1$.

\begin{acknowledgements}
    This work was supported by the EPSRC Grant T23725/01.
\end{acknowledgements}
\newcommand{\Name}[1]{{\scshape #1},}
\newcommand{\Review}[1]{{\itshape #1},}
\newcommand{\Vol}[1]{{\bfseries #1}}
\newcommand{\Year}[1]{(#1)}
\newcommand{\Page}[1]{{\normalfont #1}}
\newcommand{\REVIEW}[4]{\Review{#1} \Vol{#2} \Year{#3} \Page{#4}}
\newcommand{\etal}{\unskip\ \emph{et al.}}
\newcommand{\Book}[1]{ {\itshape #1}}
\def\and{{\normalfont{and}} }

\end{document}